\documentclass[twocolumn]{aastex63}
\usepackage{graphics,epsf}
\usepackage{amsmath}                % American Mathematical Society package
\usepackage{amsfonts}               % American Mathematical Society fonts
\usepackage{amssymb}                % American Mathematical Society symbol
\usepackage{epsfig}                 % EPS figures
\usepackage{appendix}
\usepackage{graphicx}
\usepackage{float}
\usepackage{color}
\usepackage{multirow}
\usepackage{colortbl}
\usepackage[para,online,flushleft]{threeparttable}

\newcommand{\kms}{{~\rm km\; s^{-1}}}

\newcommand{\cm}{{~\rm cm}}

\newcommand{\s}{{~\rm s}}
\newcommand{\h}{{~\rm h}}

\newcommand{\g}{{~\rm g}}

\newcommand{\erg}{{~\rm erg}}
\newcommand{\yr}{{~\rm yr}}

\newcommand{\eV}{{~\rm eV}}

\begin{document}

\title{Simulating the outcome of binary neutron star merger in common envelope jets supernovae}

%\correspondingauthor{Ealeal Bear, Noam Soker}
%\email{soker@physics.technion.ac.il}

\author{Muhammad Akashi}
\affiliation{Department of Physics, Technion, Haifa, 3200003, Israel; akashi@physics.technion.ac.il; soker@physics.technion.ac.il}
\affiliation{Kinneret College on the Sea of Galilee, Samakh 15132, Israel}

\author{Noam Soker}
\affiliation{Department of Physics, Technion, Haifa, 3200003, Israel; akashi@physics.technion.ac.il; soker@physics.technion.ac.il}
\affiliation{Guangdong Technion Israel Institute of Technology, Guangdong Province, Shantou 515069, China}

\begin{abstract}
We simulate the influence of the energy that the merger process of two neutron stars (NSs) releases inside a red supergiant (RSG) star on the RSG envelope inner to the merger location. In the triple star common envelope evolution (CEE) that we consider a tight binary system of two NSs spirals-in inside an RSG envelope and because of mass accretion and dynamical friction the two NS merge. We deposit merger-explosion energies of $3 \times 10^{50}$ and $10^{51} \erg$ at distances of $25 R_\odot$ and $50 R_\odot$ from the center of the RSG, and with the three-dimensional hydrodynamical code FLASH we follow the evolution of the RSG envelope in inner regions. For the parameters we explore we find that more than 90 per cent of the RSG envelope mass inner to the merger site stays bound to the RSG.  NSs that experience a CEE are likely to accrete RSG envelope mass through an accretion disk that launches jets. These jets power a luminous transient event, a common envelope jets supernova (CEJSN). The merger process adds to the CEJSN energy.  Our finding implies that the interaction of the merger product, a massive NS or a BH, with the envelope can continue to release more energy, both by further in-spiral and by mass accretion by the merger product. Massive RSG envelopes can force the merger product to spiral-in into the core of the RSG, leading to an even more energetic CEJSN. 
\end{abstract}


% ==========================================================
\section{Introduction} 
\label{sec:intro}
% ==========================================================

When a neutron star (NS) or a black hole (BH), hereafter NS/BH, experience a common envelope evolution (CEE) with a red supergiant (RSG) star, the NS/BH spirals in inside the RSG envelope and can accrete mass at a very high rate of $\dot M_{\rm acc} \ga 10^{-3} M_\odot \yr^{-1}$ due to neutrino cooling \citep{HouckChevalier1991, Chevalier1993, Chevalier2012}. The density gradient in the envelope results in a net angular momentum of the mass that the NS/BH accretes, and the accretion proceeds via an accretion disk around the NS/BH is very likely to launch jets (e.g.,  \citealt{ArmitageLivio2000, Papishetal2015, SokerGilkis2018, LopezCamaraetal2019, LopezCamaraetal2020MN, Hilleletal2021}). The jets remove high-energy gas from the accreting object vicinity, reducing the pressure in the NS/BH vicinity and further facilitate the accretion process  (e.g, \citealt{Shiberetal2016, Staffetal2016, Chamandyetal2018}). A BH accretor can as well directly accrete high-energy gas (e.g., \citealt{Pophametal1999}). 
 
At early phases of the CEE the jets inflate the envelope, eject mass, and substantially increase the luminosity of the system. At later phases the jets explode the envelope and if the NS/BH enter the core the jets explode the entire star in a common envelope jets supernova (CEJSN) event (e.g., \citealt{Gilkisetal2019, GrichenerSoker2019a, SokeretalGG2019, Schroderetal2020, GrichenerSoker2021}). A different type of explosion where an old NS accretes mass within a RSG is the explosion of a massive Thorne–Zytkow object as \cite{Moriya2018} suggests (we note that Thorne–Zytkow objects might form through a CEE with a WD rather than a NS; \citealt{Ablimitetal2021}). In case the NS/BH does not enter the core, and in particular if the NS/BH enters and exits the RSG envelope, the transient event is a CEJSN impostor (e.g., \citealt{Gilkisetal2019, Schreieretal2021}). The inflation of the envelope and removal of large amounts of envelope mass by the jets can increase the CEE parameter to be $\alpha_{\rm CE} > 1$, as a number of CEE scenarios assume (e.g. \citealt{Fragosetal2019, BroekgaardenBerger2021, Fragioneetal2021, Garciaetal2021, Zevinetal2021}). The jets of CEJSNe, probably with a NS accretor, might be a site of r-process nucleosynthesis \citep{Papishetal2015, GrichenerSoker2019a, GrichenerSoker2019b}. In the case of a BH accretor the jets might be the source of very-high-energy ($\approx 10^{15} \eV$) neutrinos \citep{GrichenerSoker2021}. 

There are different types of CEJSNe (and impostors) in binary systems, some of which might account for puzzling transient events (e.g., \citealt{Thoneetal2011, SokerGilkis2018, SokeretalGG2019}). Recent studies, which are motivated also by the large fraction of massive triple stellar systems (e.g., \citealt{Sanaetal2014, MoeDiStefano2017}), have added a rich variety of evolutionary routes of CEJSNe in triple stellar systems (e.g., \citealt{Schreieretal2021, Soker20212CEJSNe, Soker2021NSNS}). 
\cite{Soker2021NSNS} explores several evolutionary routes of a triple-star scenario where a tight NS-NS binary system enters a CEE with a RSG. These evolutionary routes include routes where the NS-NS tight binary system merges inside the RSG envelope or later in its core. We here consider the case where the NS-NS merger process takes place in the envelope and examine the envelope mass that stays bound to the system after the NS-NS merger.  

In section \ref{sec:Evolution} we describe the evolutionary route (one out of several that \citealt{Soker2021NSNS} studies). We describe the numerical procedure in section \ref{sec:Numerics} and our main results in section \ref{sec:Results}. We summarise our findings in section \ref{sec:Summary}. 

% ==========================================================
\section{The evolutionary route} 
\label{sec:Evolution}
% ===========================================================

\begin{figure*}[t]
 %\centering
%  \vskip -3.00 cm
%\hskip -0.90 cm
\center
\includegraphics[trim=0.0cm 7.6cm 0.0cm 2.3cm ,clip, scale=0.90]{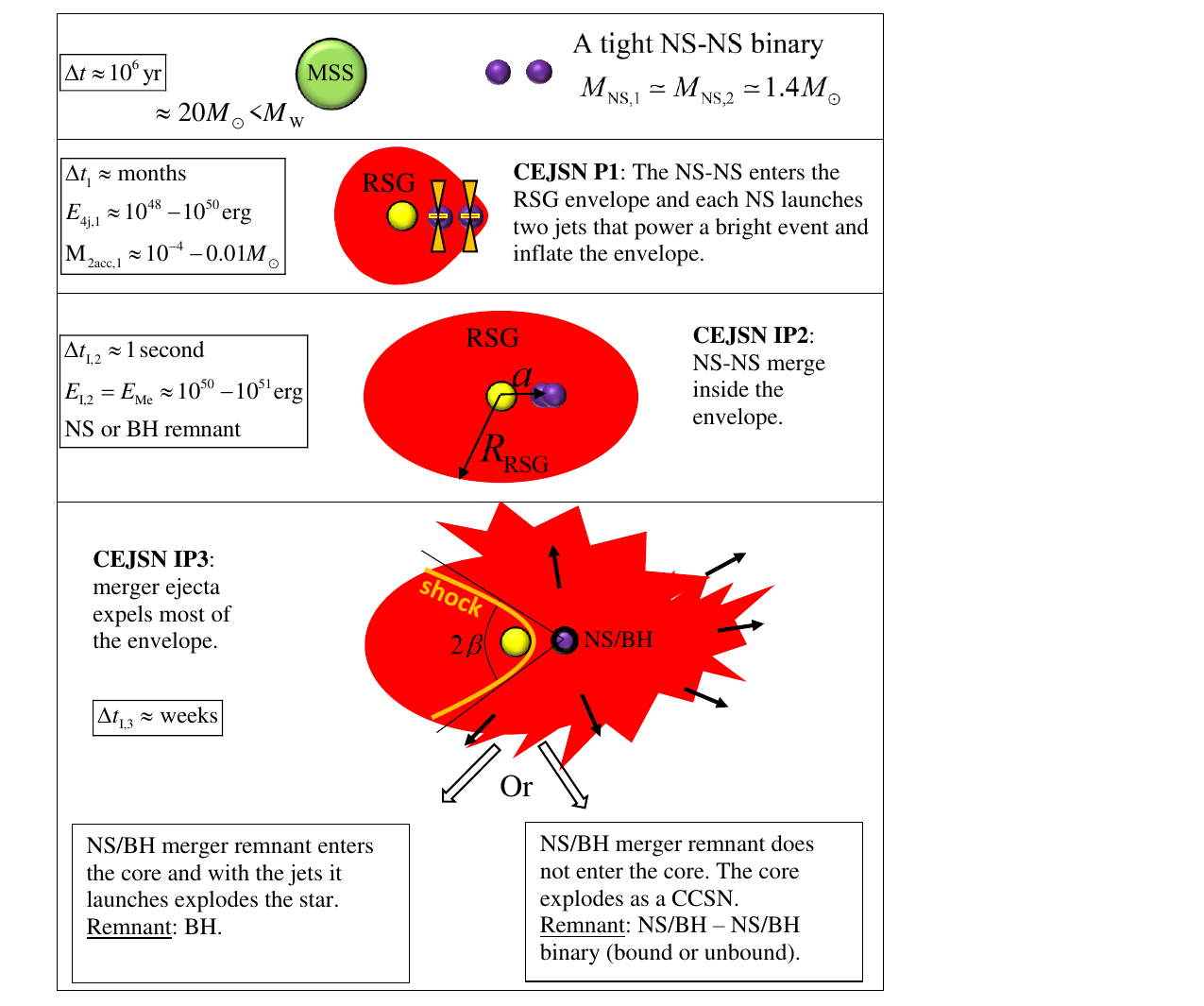}
% \includegraphics[]{CCSNIaFigure.pdf}\\
% \vskip -6.00 cm
\caption{A schematic diagram of the evolutionary route where the tight NS-NS binary system merges inside the envelope (adopted from figure 1 of \citealt{Soker2021NSNS}). The time $\Delta t \approx 10^6 \yr$ in the first panel is the time from the formation of the tight NS binary system until the onset of the triple star CEE. The merger inside the envelope (phase IP2) explosively releases an energy of $E_{\rm mer} \approx 10^{50} - 10^{51} \erg$ that unbinds most of the envelope (phase IP3). We mark two ends of the evolutionary route at the bottom of the figure (for more details of these two possible outcomes see \citealt{Soker2021NSNS}). 
We simulate the interaction of the merger ejecta with the inner envelope for 10 hours. The time scale of $\Delta_{\rm I,3} \approx {\rm weeks}$ refers to the time the shock propagates through the entire inflated envelope, to about $\approx 10^4 R_\odot$. 
Abbreviation: BH: black hole; CCSN: core collapse supernova; MSS: main sequence star; NS: neutron star; RSG: red supergiant. 
}
 \label{fig:SchematicScenario}
 \end{figure*}

In Fig. \ref{fig:SchematicScenario} we schematically describe the relevant phases of the evolutionary route that we study here (more details are in Fig. 1 of \citealt{Soker2021NSNS}). The scenario is of a tight NS-NS binary system that enters the envelope of a RSG star. One of the several outcomes of a tight binary system that enters a CEE is the merger of the two components of the tight binary system, (e.g., \citealt{SabachSoker2015, Hilleletal2017, ComerfordIzzard2020, GlanzPerets2021,  Soker20212CEJSNe} for studies that did not consider NS-NS binary systems). Due to tidal interaction and mass accretion the two NSs merge inside the envelope. The NS-NS merger remnant is a very massive NS or a BH (NS/BH). Before merger the NSs accrete mass and launch jets, a process that inflates the envelope, ejects mass, and energises the early light curve of the CEJSN event (or a CEJSN impostor event if the NS/BH does not enter the core at the end). 
%FFFFFFFFFFFFFFFFFFFFFFFFFFFFFFFFFFFFFFFFF
  %[ht]
 % %FFFFFFFFFFFFFFFFFFFFFFFFFFFFFFFFFFFFFF 

In this study we simulate only the effect of the  explosive energy deposition by the merging NSs on the RSG envelope. According to analysis of the NS merger (NSM) event GW170817 and numerical simulations (e.g., \citealt{Hajelaetal2021} and references therein) the kinetic energy of the NS-NS merger ejecta is $E_{\rm mer} \approx 10^{50} - 10^{51} \erg$ (e.g., \citealt{Trojaetal2020}), the ejecta mass from the NS merger process is $M_{\rm mer} \simeq 0.06 M_\odot$ (the equatorial ejecta is in the bulk range of $M_{\rm mer} \simeq 0.01-0.1 M_\odot$, e.g., \citealt{Metzger2019}), and the ejecta velocities are in the range of $v_{\rm mer} \simeq 0.1-0.3 c$ (e.g., \citealt{Metzger2019}). 

The exact geometry of the NS-NS merger ejecta is uncertain. The NS-NS merger process ejects material both in the polar directions and in the equatorial plane of the two NSs (e.g., \citealt{Metzger2019}). In the scenario that we study here ejecta in all directions collide with the gas in the RSG envelope and form a hot bubble around the merger site. We follow \citealt{Soker2021NSNS} and take the merger-explosion  bubble to be spherical. 

\cite{Soker2021NSNS} very crudely estimates that the NS-NS merger process ejects most of the envelope mass, and that for RSG envelope masses of $M_{\rm RSG,e} \ga 10 M_\odot$ an envelope mass of $M_{\rm PM,e} \approx 0.5-5 M_\odot$ stays bound. 
In estimating the bound mass \cite{Soker2021NSNS} assumed that most of the merger energy escapes to directions away from the direction of the core with respect to the merger site because the envelope densities are low in those directions. Therefore, he assumed, the interaction of the merger ejecta with the RSG envelope within small angles $\la \beta$ to the core of the RSG is momentum conserving. From this assumption he calculated the outflow velocity of the RSG envelope that the merger ejecta accelerate out, and estimated the envelope mass fraction that does not reach the escape velocity. 

We set the goal of examining this bound mass.  

% ==========================================================
\section{Numerical procedure} 
\label{sec:Numerics}
% ===========================================================

% ===========================
\subsection{The RSG model} 
\label{subsec:RSGmodel}
% ===========================

\begin{figure}[t]
	%\centering
\includegraphics[trim=1.8cm 12.0cm -1.0cm 4.5cm ,clip, scale=0.50]{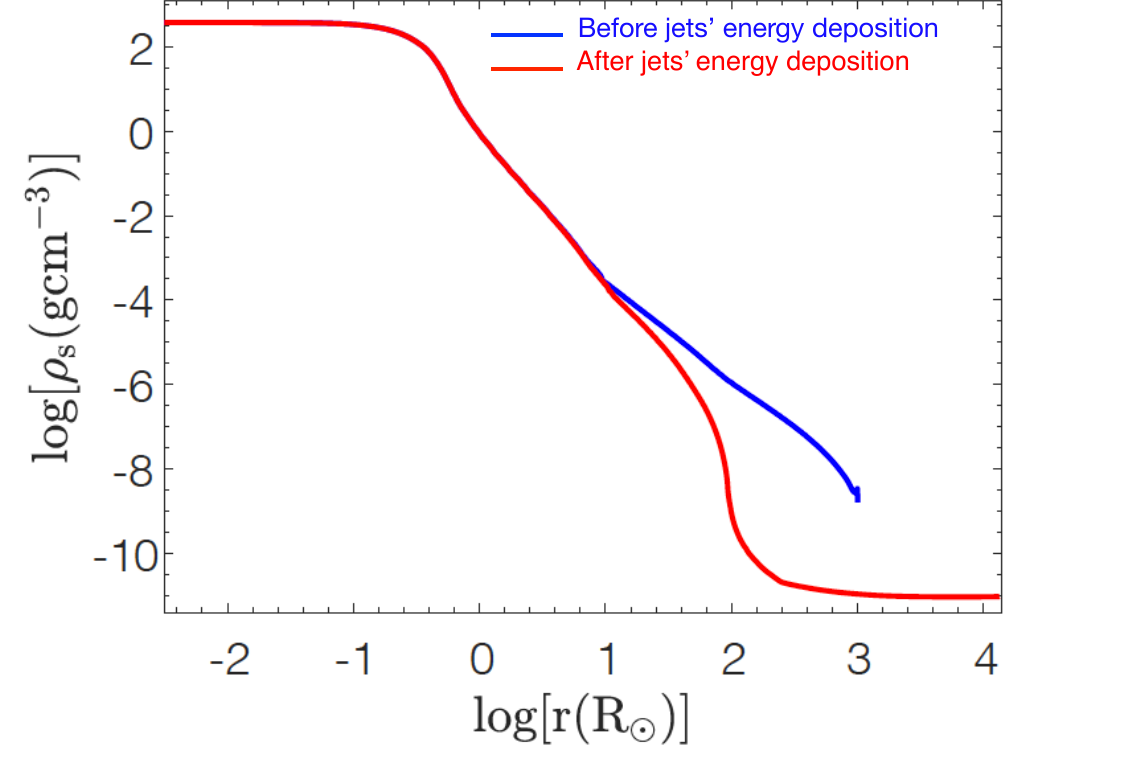} \\
\caption{Density profile of the RSG when it reaches a radius of $R=1000R_{\rm \odot}$  (blue curve) and at the end of three-years energy deposition to mimic jets from a NS/BH companion.The red profile is the initial (at $t=0$) density profile of our RSG model.  From \cite{GrichenerSoker2021}.  }
\label{fig:DensityProfile}
\end{figure}

We import a spherical RSG model with an inflated envelope as \cite{GrichenerSoker2021} built with the stellar evolution code \textsc{mesa} (Modules for Experiments in Stellar Astrophysics; \citealt{Paxtonetal2011, Paxtonetal2013, Paxtonetal2015, Paxtonetal2018, Paxtonetal2019}). 
The stellar model is of  a non-rotating star of initial mass $M_{\rm 2,ZAMS}=30M_{\rm \odot}$ and metalicity of $Z=0.02$. 
When the star becomes a RSG with a radius of $R_{\rm 2} = 1000 R_{\rm \odot}$ \cite{GrichenerSoker2021} assume that it swallows a NS/BH that launches jets that inflate the envelope. At this time the mass of the RSG star is $M_{\rm 2}=20M_{\rm \odot}$ because of wind mass loss during its evolution to the RSG phase. After energy deposition in the outer parts of the envelope the radius of the RSG is $R_{\rm 2,RSG}=1.3\times 10^{4} R_{\rm \odot}$ (for more details and justifications see \citealt{GrichenerSoker2021}). 
We present the density profiles before energy deposition and after energy deposition in Fig. \ref{fig:DensityProfile}. 
% FFFFFFFFFFFFFFFFFFFFFFFFFFFFFFFFFFFFFFFF
% FFFFFFFFFFFFFFFFFFFFFFFFFFFFFFFFFFFFFFFF
 
\begin{figure}[t]
	%\centering
\includegraphics[trim=0.6cm 7.5cm 0.cm 7.0cm ,clip, scale=0.45]{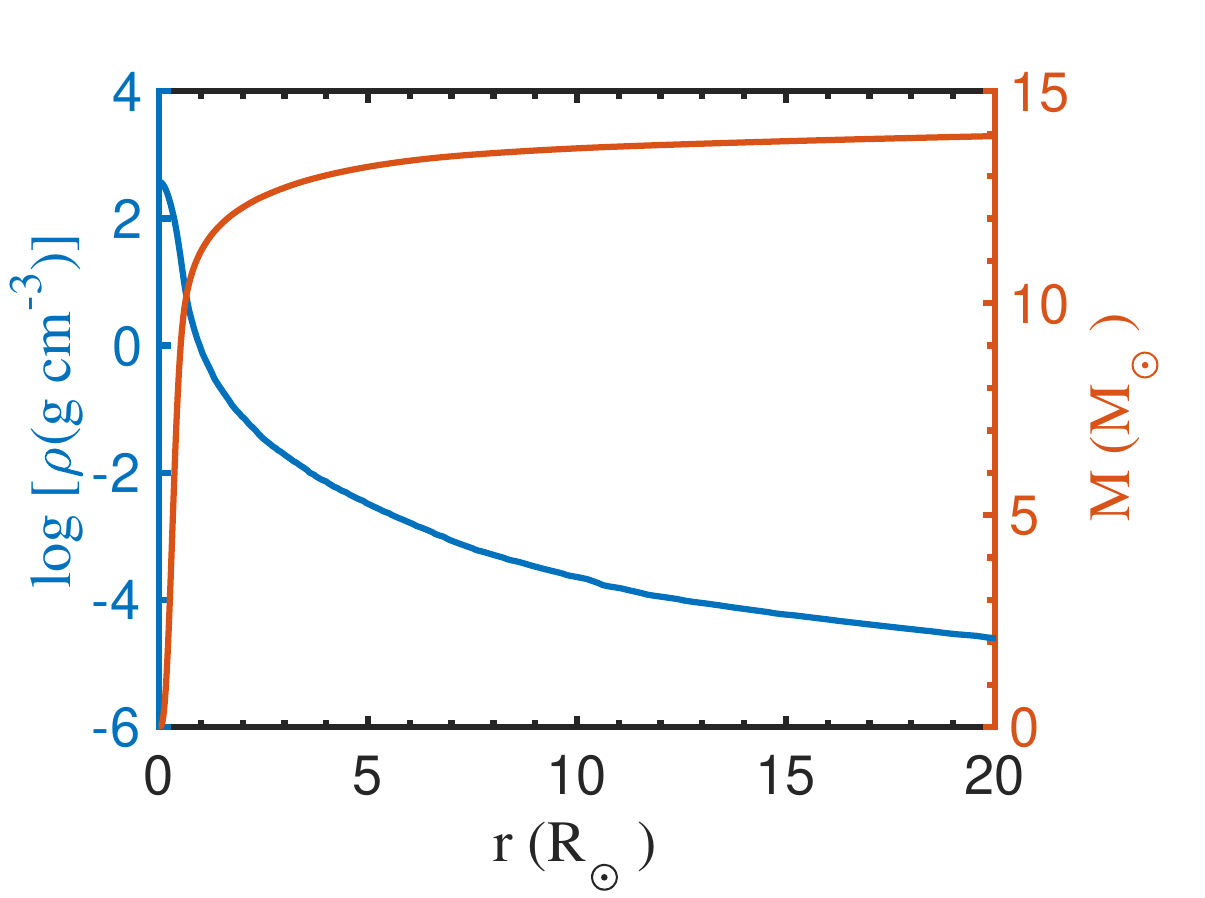} \\

\caption{Density and mass profiles in the inner $R=20R_{\rm \odot}$  profile of the RSG model that we start with (corresponding to the inflated model of Fig. \ref{fig:DensityProfile}.) }
	\label{fig:Dens_mass}
\end{figure}

\cite{GrichenerSoker2021} deposited the energy that mimics the jets that the NS/BH launches to the outer zone of the envelope. This inflates the envelope to huge dimensions, but does not change much the density profile in the inner envelope $r \la 10 R_\odot$. As we will show, a large fraction of the envelope mass that stays bound after the NS-NS merger comes from this inner zone of the envelope. For that, in Fig. \ref{fig:Dens_mass} we present the density and mass profile in the inner $20 R_\odot$. 
% FFFFFFFFFFFFFFFFFFFFFFFFFFFFFFFFFFFFFFF
% FFFFFFFFFFFFFFFFFFFFFFFFFFFFFFFFFFFFFFF

% ===========================
\subsection{Three-dimensional hydrodynamical procedure}
% ===========================

We use version 4.6.2 of the adaptive-mesh refinement (AMR) hydrodynamical FLASH code \citep{Fryxell2000} in three dimension (3D). 
 
As the gas is optically thick in most regions of the RSG star, we turn off radiative cooling at any gas temperature.
The equation of state includes both radiation pressure and gas pressure with an adiabatic index of $\gamma=5/3$, due both to ions and electrons, i.e., $P_{\rm tot} = P_{\rm rad} + P_{\rm ion}+P_{\rm elec}$. 
  
The composition that we use in FLASH is fully ionised pure hydrogen. Because of the 3D Cartesian (rather than spherical) numerical grid and the different composition from MESA, the 3D stellar model in FLASH is not completely in hydrostatic equilibrium. However, we found that the time it takes the star to substantially deviate from its original structure is longer than the time it takes the shock of the merger process to propagate through the computational grid. 
  
The grid is AMR 3D Cartesian with outflow boundary conditions. We use resolution of 7 refinement levels, which means that the ratio between the grid size to the smallest cell size $\Delta_{\rm cell,m}$ is $2^{9}$. 
We simulate the whole space (the two sides of the equatorial plane). The full size of the Cartesian numerical grid is either $(200R_{\rm \odot})^3$, i.e., $(L_x,L_y,L_z) = \pm 100 R_{\rm \odot}$, or $(120 R_{\rm \odot})^3$ in the high resolution (HR) simulations.
In Table \ref{Table:cases} we list the different simulations with their properties. 
% TTTTTTTTTTTTTTTTTTTTTTTTTTTTTTTTTTTT
\begin{table}%[]
\centering
\begin{tabular}{|c|c|c|c|c|c|c|c|}
\hline

Simulation &  $R_{\rm inert}$ &  $\Delta_{\rm cell,m}$  & $X_{\rm mer}$ & $E_{\rm exp}$ \\ 
           &  $R_{\rm \odot}$ &  $R_{\rm \odot}$  & $R_{\rm \odot}$  & $\erg$ \\
           \hline 
S1         & $2$ &0.39 & $50$ & $10^{51}$  \\ \hline
S2 (HR)    & $2$ & 0.234& $50$ & $10^{51}$  \\ \hline
S3 (HR)    & $2$ &0.234 & $50$ & $3 \times 10^{50}$ \\ \hline
S4 (HR)    & $2$ &0.234 & $25$ & $10^{51}$  \\ \hline
S5 (HR)    & $2$ & 0.234& $25$ & $3 \times 10^{50}$ \\ \hline
S6         & $3$ & 0.39& $50$ & $10^{51}$  \\ \hline
S7         & $4$ &0.39 & $50$ & $10^{51}$  \\ \hline
S8         & $5$ & 0.39& $50$ & $10^{51}$  \\ \hline

\end{tabular}
\caption{Summary of the 8 simulations we present in the paper. The columns list, from left to right and for each simulation, its name, the radius of the inert core $R_{\rm inert}$, the minimum cell size in the numerical grid $\Delta_{\rm cell,m}$, the location of the merger relative to the center of the RSG $X_{\rm mer}$, and the merger explosion energy $E_{\rm exp}$.}
\label{Table:cases}
\end{table}
% TTTTTTTTTTTTTTTTTTTTTTTTTTTTTTTTTTTT
  
At $t=0$ we place the RSG stellar model from  \cite{GrichenerSoker2021} at the center of the grid $(0,0,0)$, as we described above in section \ref{subsec:RSGmodel}. Our grid contains a small part of the RSG model. To avoid numerical instabilities and to allow reasonable time steps, during the entire simulation we fix a spherical volume at the center, the `inert core', of radius $R_{\rm inert}$ inside which we allow no changes of the hydrodynamical variables, e.g., the velocity inside the inert core is zero. 
We do not include self-gravity of the envelope nor the gravity of the merger product. We do include the gravity of the core of mass $10.1 M_\odot$ as a point mass at the center of the grid and that is constant in time. 

The non-inclusion of the self gravity of the envelope above mass coordinate of $10.1 M_\odot$ has a little effect for the following reason. Most of the mass inner to the radii where we set the merger explosion is of the core, what we take at $10.1 M_\odot$. The exact gravitational energy of the entire envelope above $10.1 M_\odot$ is  $-1.14 \times 10^{50} \erg$. 
The gravitational energy of the envelope mass above  $10.1 M_\odot$ when we include only the gravity of the $10.1 M_\odot$ core is $-1.02 \times 10^{50} \erg$.  
The difference of $1.2 \times 10^{49} \erg$ between these two values is much smaller than the merger explosion energy. Therefore, we estimate that including the gravity of the NSs and of the mass of the envelope above $10.1 M_\odot$ will not change much. It will increase a little the amount of bound mass.

We set a NS-NS merger-explosion at $(x,y,z)=(X_{\rm mer},0,0)$. We mimic the merger explosion by inserting the explosion energy $E_{\rm exp}$ at $t=0$ inside a spherical volume with a radius of $R_{\rm mer}=5R_\odot$ centred at the merger site. At $t=0$ we set this explosion volume to have a constant density with a total mass of $M_{\rm mer} = 0.05M_{\odot}$ for the $E_{\rm exp}=10^{51} \erg$ simulations and $M_{\rm mer} = 0.015M_{\odot}$ for the $E_{\rm exp}=3\times 10^{50} \erg$ simulations, and with a linear radial velocity field as measures from the center of explosion (the merger site), i.e., ${\vec v}(r_{\rm mer})=({{\vec r}_{\rm mer}}/5R_{\rm \odot})v_{\rm m}$, where $v_m = 57900 \kms $ and $r_{\rm mer}$ is the distance from the merger site.
The initial kinetic energy inside the explosion sphere equals the explosion energy, and the initial temperature is very low.

\label{subsec:3Dhydro}
% ===========================

% =====================================  

% ==========================================================
\section{Results} 
\label{sec:Results}
% ===========================================================

% =================================
\subsection{The flow structure} 
\label{subsec:Results}
% =================================

We present the flow structure as function of time for simulation S2 (HR) that has an inert core of $R_{\rm inert} = 2 R_\odot$. We found that as we decrease the inert core we need a higher resolution. Our computer resources limit us to use inert cores that are not much smaller than $R_{\rm inert} = 2 R_\odot$. In section \ref{subsec:BoundMass} we present the influence of the radius of the inert core and of the numerical resolution on the results. 
 
\begin{figure*}%[]
	%\centering
\includegraphics[trim=0.1cm 6.0cm 0.0cm 0.0cm ,clip, scale=0.91]{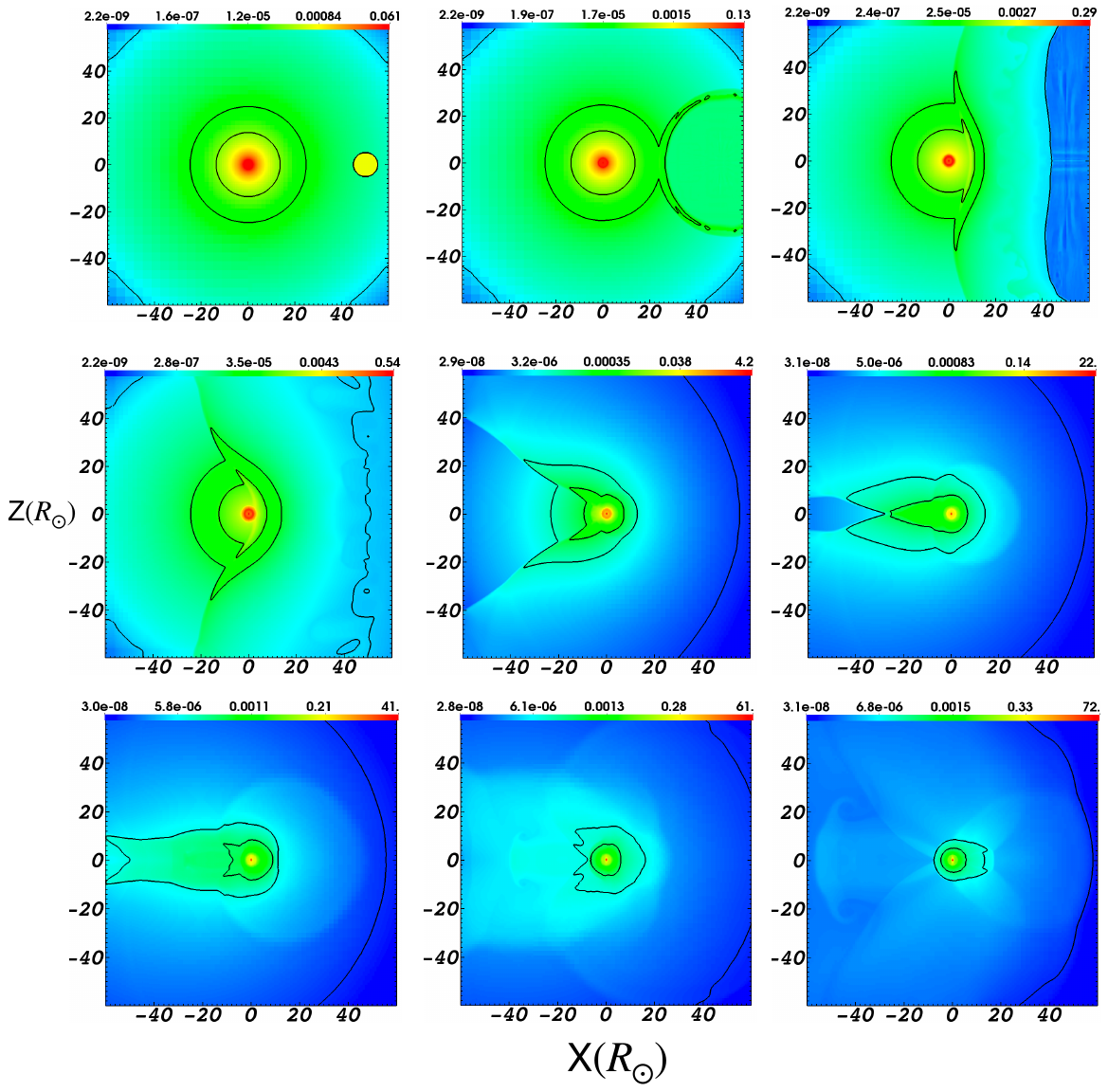} \\
\caption{Density maps at nine times of 0, 0.1, 0.5, 1, 2, 3, 4, 6, and $10 \h$, from upper left to lower right. All panels are square with sizes of $120 R_{\rm \odot}$. The colors depict the density values in $\g \cm^{-3}$ according to the color bar on the top of each panel. We also plot contours for three densities of $10^{-7}$, $10^{-5}$, and $10^{-4} \g \cm^{-3}$.  The location of the merger site is at $X_{\rm mer}=50R_{\rm \odot}$, and the center of the RSG is at the center $(0,0,0)$. }
	\label{fig:dens_cont}
\end{figure*}

\begin{figure*}%[]
	%\centering
\includegraphics[trim=0.1cm 4.81cm 0.0cm 0.0cm ,clip, scale=0.87]{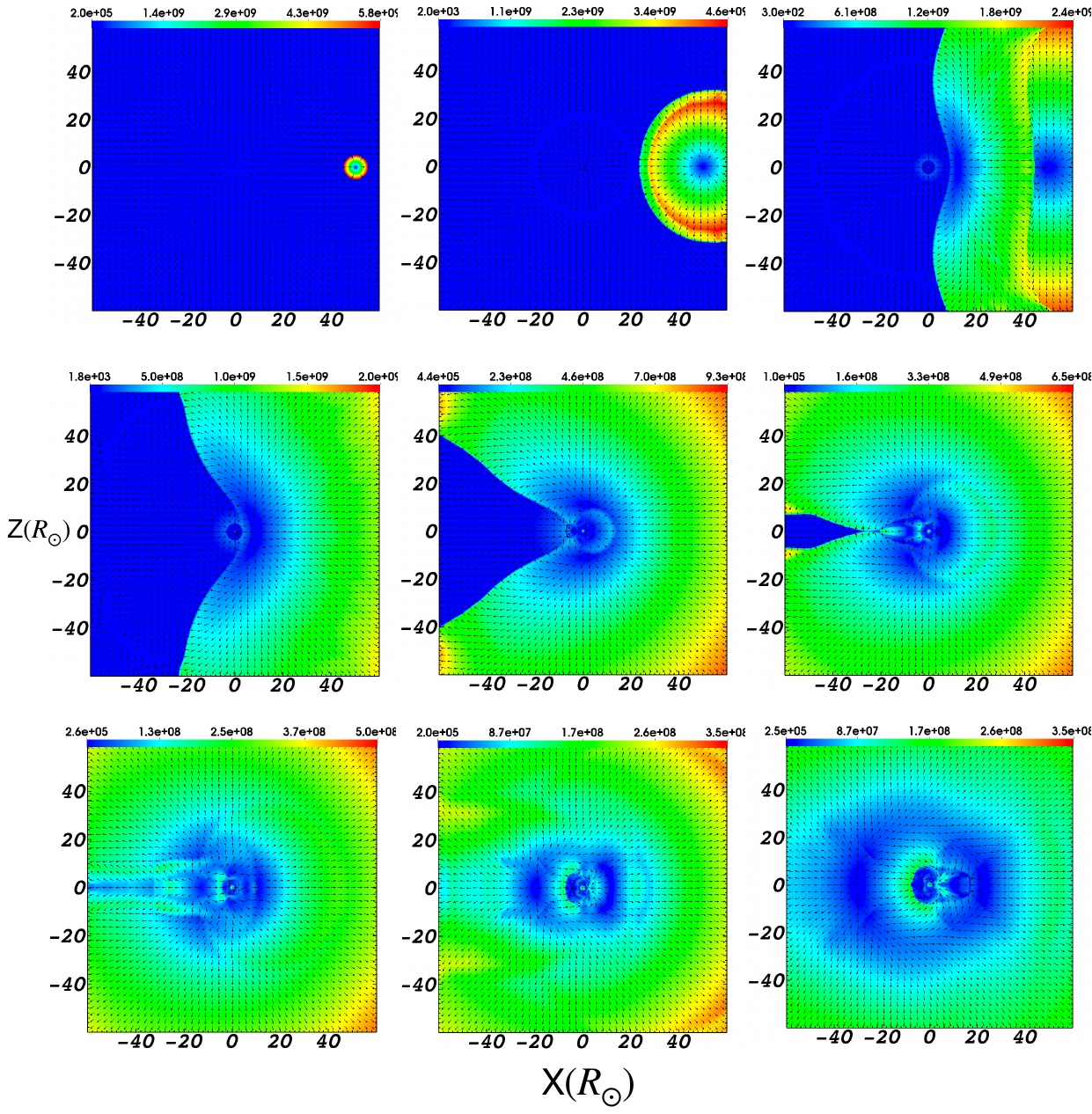} \\

\caption{Velocity maps at nine times as in Fig. \ref{fig:dens_cont}. The colors depict the velocity values according to the color bars on the top of each panel in $\cm \s^{-1}$ (in the first four panels the largest power in the color bars is 9 and in the last five panels the largest powers is 8). The arrows show the velocity directions.  }
	\label{fig:vmag_arrows}
\end{figure*}

\begin{figure*}%[]
	%\centering
\includegraphics[trim=0.1cm 4.53cm 0.cm 0.0cm ,clip, scale=0.87]{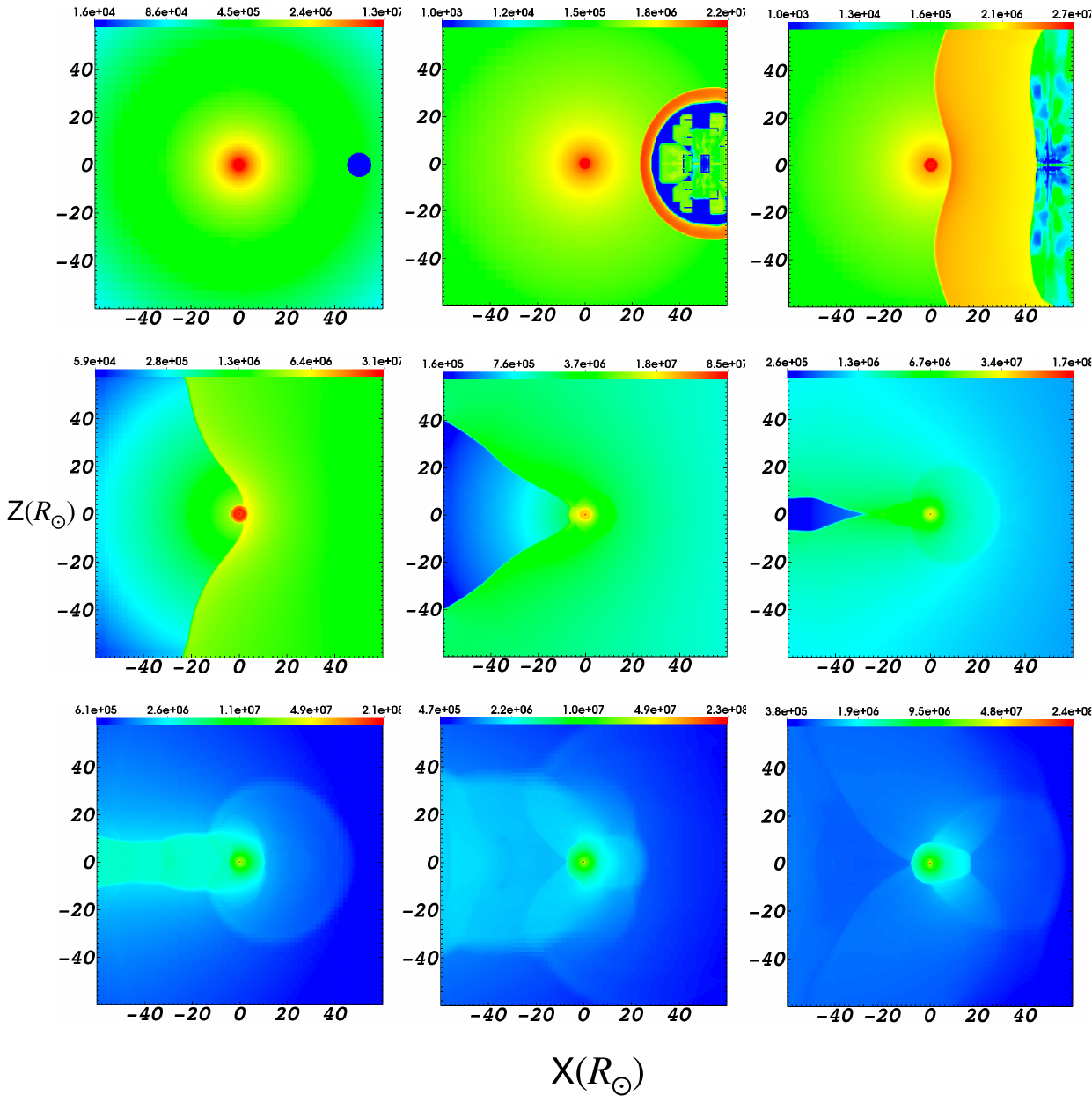} \\

\caption{Temperature maps at nine times as in Fig. \ref{fig:dens_cont}. The colors depict the temperature values in K according to the color bars on the top of each panel (in the first five panels the largest power in the color bars is 7 and in the last four panels the largest powers is 8).}
	\label{fig:temp}
\end{figure*}

Because we neglect envelope rotation and deformation from spherical symmetry (but we do include envelope inflation) due to the pre-merger spiralling-in of the NS binary and the jets that the NSs might launch, our computation has an axial symmetry around the line from the core to the merger site. We therefore present the results only in the $(x,z)$ plane that contains the center of the core and the merger site. In the upper left panel of Fig.  \ref{fig:dens_cont} we present the density at $t=0$. The core is at the center of the red sphere at $(0,0)$ and the merger site is at the center of the yellow sphere at $(x,z)=(50R_\odot, 0)$. The next eight density panels present the evolution until $t= 10 \h$. 
Similarly, in Fig. \ref{fig:vmag_arrows} we present the velocity maps at the same nine times. The colors represent the magnitude of the velocity and the arrows represent the directions. In Fig. \ref{fig:temp} we present the temperature maps at the same nine times. 
 There are square patches around the explosion site and inside the bubble that the explosion inflates in the two upper right panels of Fig. \ref{fig:temp}. We attribute these to numerical perturbations from the right grid boundary that develop as the very high pressure shock wave reaches the right grid boundary. These numerical perturbations propagate only inside the bubble that the merger-explosion inflates that contains a small fraction of the explosion energy, and they do not affect much the shock wave that runs into the RSG envelope (the red arc in the middle-upper panel and the orange strip in the right-upper panel).
% FFFFFFFFFFFFFFFFFFFFFFFFFFFFFFFFFFFFFFF
% FFFFFFFFFFFFFFFFFFFFFFFFFFFFFFFFFFFFFFF
% FFFFFFFFFFFFFFFFFFFFFFFFFFFFFFFFFFFFFFF    
% FFFFFFFFFFFFFFFFFFFFFFFFFFFFFFFFFFFFFFF
% FFFFFFFFFFFFFFFFFFFFFFFFFFFFFFFFFFFFFFF
% FFFFFFFFFFFFFFFFFFFFFFFFFFFFFFFFFFFFFFF

The shock wave that the merger-explosion excites reaches the inert sphere at about $t \simeq 1 \h$ as we can see in the middle-left panel of each of the three figures. At about $t \simeq 3 \h$ the shock closes on itself behind the core at about $x \simeq -30 R_\odot$ (middle-right panel in the three figures).
At $t \ga 3 \h$ a complicated flow develops around the inert core. We do not trust our simulations beyond about $t \simeq 10 \h$, and we terminate the simulations. The reason is that at about $t \simeq 10 \h$ numerical instabilities near the inert radius start to grow because we do not have a self-consistent treatment of the stellar structure of the bound mass and because of the finite resolution of our Cartesian grid.  We show below and in section \ref{subsec:BoundMass} that by that time the bound mass reaches a more or less steady state. 
  
At $t=0$ the $\rho = 10^{-5} \g \cm^{-3}$ contour is at a radius of $r=25 R_\odot$. At $t=10 \h$ the same density contour closes a non-spherical volume, which however has only a moderate departure from spherical structure. The average radius of this volume is $\simeq 10 R_\odot$. 
For the $\rho = 10^{-4} \g \cm^{-3}$ contour the initial radius is $r=14 R_\odot$, while the final volume has a small departure from being spherical and has an average radius of $\simeq 5 R_\odot$. The smaller final radii of these contours results from the ejection of mass from the inner regions by the shock as we discuss in section \ref{subsec:BoundMass}. That the structure of the leftover envelope in the range of these two densities at $t \simeq 10 \h$ is more or less spherical after it was highly deformed by the shock (times of $\simeq 2-4 \h$), makes us confident that we can deduce the leftover envelope mass, as we do next (section \ref{subsec:BoundMass}).    
  
Because the RSG envelope is much larger than the distance of the merger-explosion site from the center the properties of the observed explosion will be as if the explosion is at the center. This transient event will be classified as a CCSN although strictly speaking it is not. The explosion will not be spherical because of the envelope deformation by the spiralling-in process and the jets that the pre-merger NSs are likely to launch (e.g., \citealt{Soker2021AM}).

% =================================
\subsection{Final bound envelope mass} 
\label{subsec:BoundMass}
% =================================
 
\begin{figure*}%[]
	%\centering
\includegraphics[trim=0.1cm 2.6cm 0.cm 3.cm ,clip, scale=0.90]{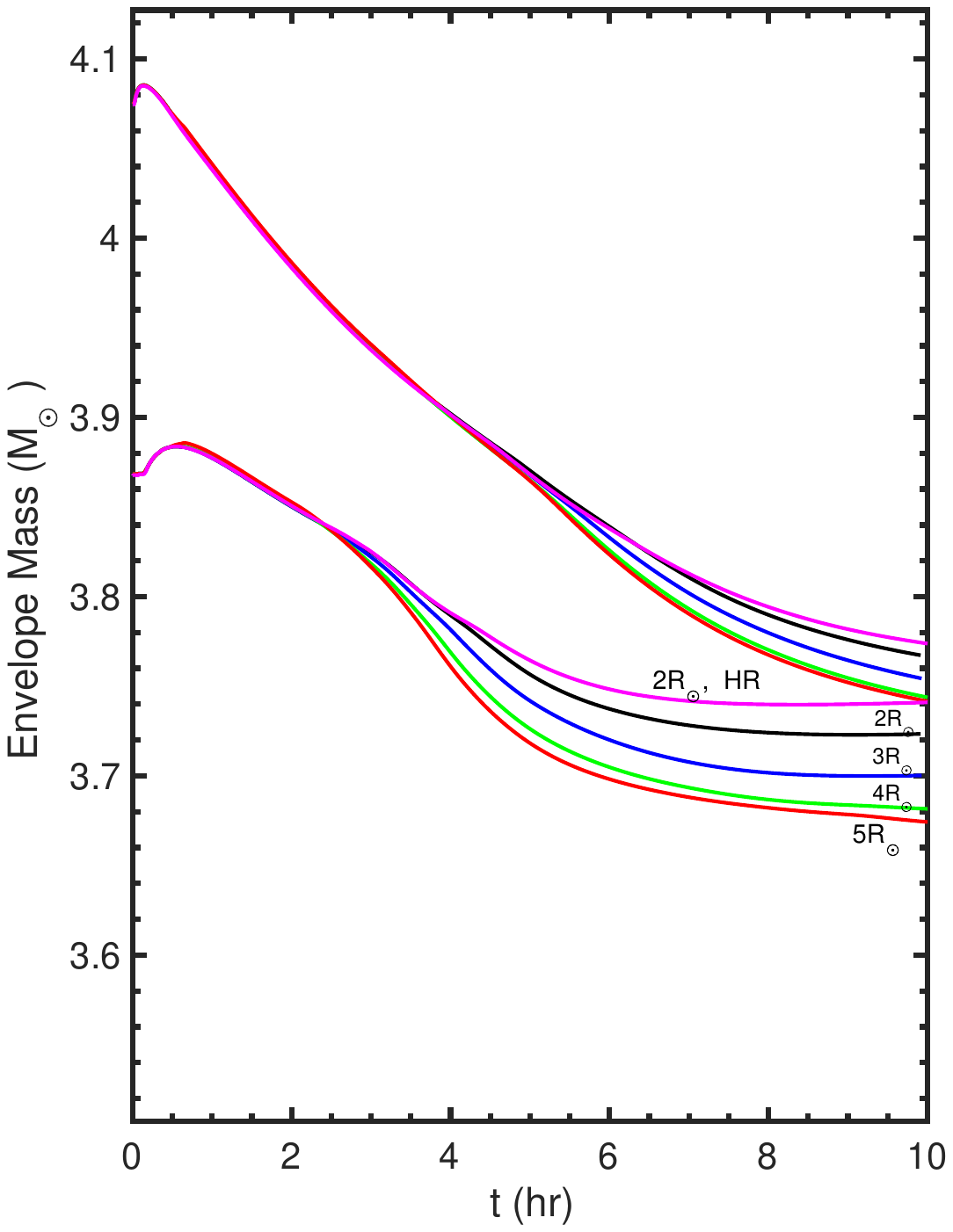} \\
\caption{Envelope mass inner to $R=20 R_\odot$ (lower group of lines) and inner to $R=40 R_\odot$ (upper group of lines) for the five simulations with $E_{\rm exp} = 10^{51} \erg$ and with $X_{\rm mer}=50 R_\odot$. The envelope mass is the mass above $r=0.64 R_\odot$ which is the radius of the $10.1 M_\odot$ helium core. The two lines for each simulation have the same color and style. We indicate the radius of the inert core near each line. HR marks high-resolution simulations. }
	\label{fig:mass2040A}
\end{figure*}

\begin{figure*}%[]
	%\centering
\includegraphics[trim=0.1cm 2.6cm 0.cm 3.cm ,clip, scale=0.90]{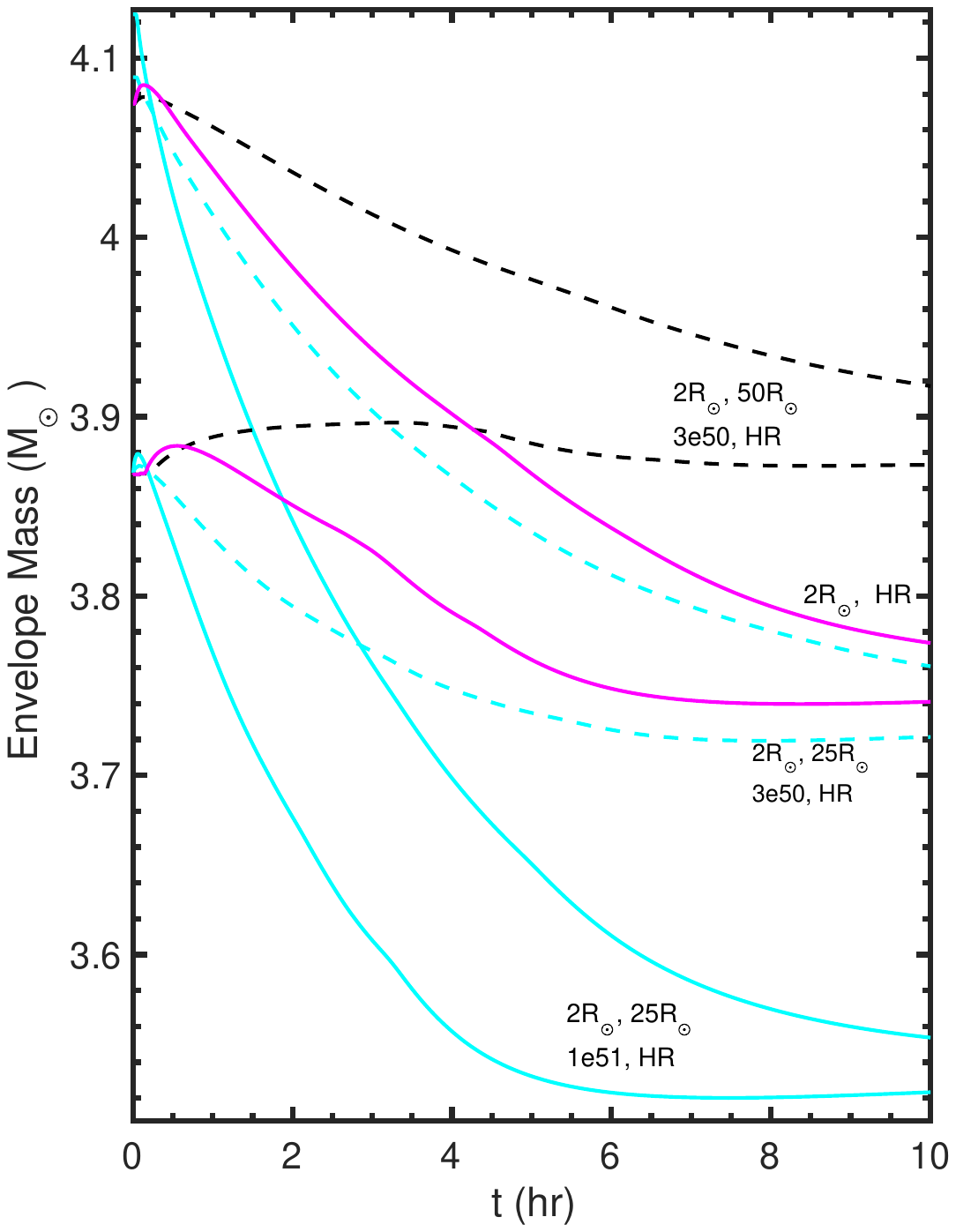} \\
\caption{Similar to Fig. \ref{fig:mass2040A} but for simulations with $R_{\rm inert}=2 R_\odot$ and different $X_{\rm mer}$ and $E_{\rm exp}$ as we mark near the lines.
 }
	\label{fig:mass2040B}
\end{figure*}

To find the leftover mass that is bound inner to the merger-explosion radius we calculate the total envelope mass inner to radii  $R=20 R_\odot$ and $R=40 R_\odot$, which we mark by $M_{20}$ and $M_{40}$, respectively. We take here the envelope mass to be the mass above $r=0.64 M_\odot$, which is the $10.1 M_\odot$ helium core mass. 
In Fig. \ref{fig:mass2040A} we present the evolution of $M_{20}$ 
(lower group of lines) and $M_{40}$ (upper group of lines) for the five simulations with $E_{\rm exp} = 10^{51} \erg$ and $X_{\rm mer}=50 R_\odot$. We mark near each line the value of the inert core. Each case has the same style and colour of lines for $M_{20}$ and $M_{40}$. In Fig. \ref{fig:mass2040B} we present the evolution of $M_{20}$ and $M_{40}$ for all cases with inert radius of $2 R_\odot$. In this figure we mark near each line the merger-explosion energy $E_{\rm exp}$ and the merger location $X_{\rm mer}$. Each case has the same style and colour of lines for $M_{20}$ and $M_{40}$.
% FFFFFFFFFFFFFFFFFFFFFFFFFFFFFFFFFFFFFFF
% FFFFFFFFFFFFFFFFFFFFFFFFFFFFFFFFFFFFFFF
% FFFFFFFFFFFFFFFFFFFFFFFFFFFFFFFFFFFFFFF

Comparing the simulations with different $R_{\rm inert}$ we learn that the leftover bound envelope mass increases as the simulations become more realistic, namely, lower $R_{\rm inert}$ and higher resolution. 
The mass in the zone from $r=2R_\odot$ to $r=5 R_\odot$ in the initial stellar model is $\Delta m_{2-5} = 0.93 M_\odot$. This is much larger than the difference between the bound masses that we find for $R_{\rm inert} = 2 R_\odot$ and for $R_{\rm inert} = 5 R_\odot$, which is $\Delta M_{\rm 2-5,bound} \la 0.06 M_\odot$. The inequality  $\Delta M_{\rm 2-5,bound} \ll \Delta m_{2-5}$ shows that our results do not depend much on the inert radius we take as long as it is $R_{\rm inert} \la 5 R_\odot$.

\begin{figure*}%[]
	%\centering
\includegraphics[trim=0.6cm 10.0cm 0.0cm 2.0cm ,clip, scale=0.85]{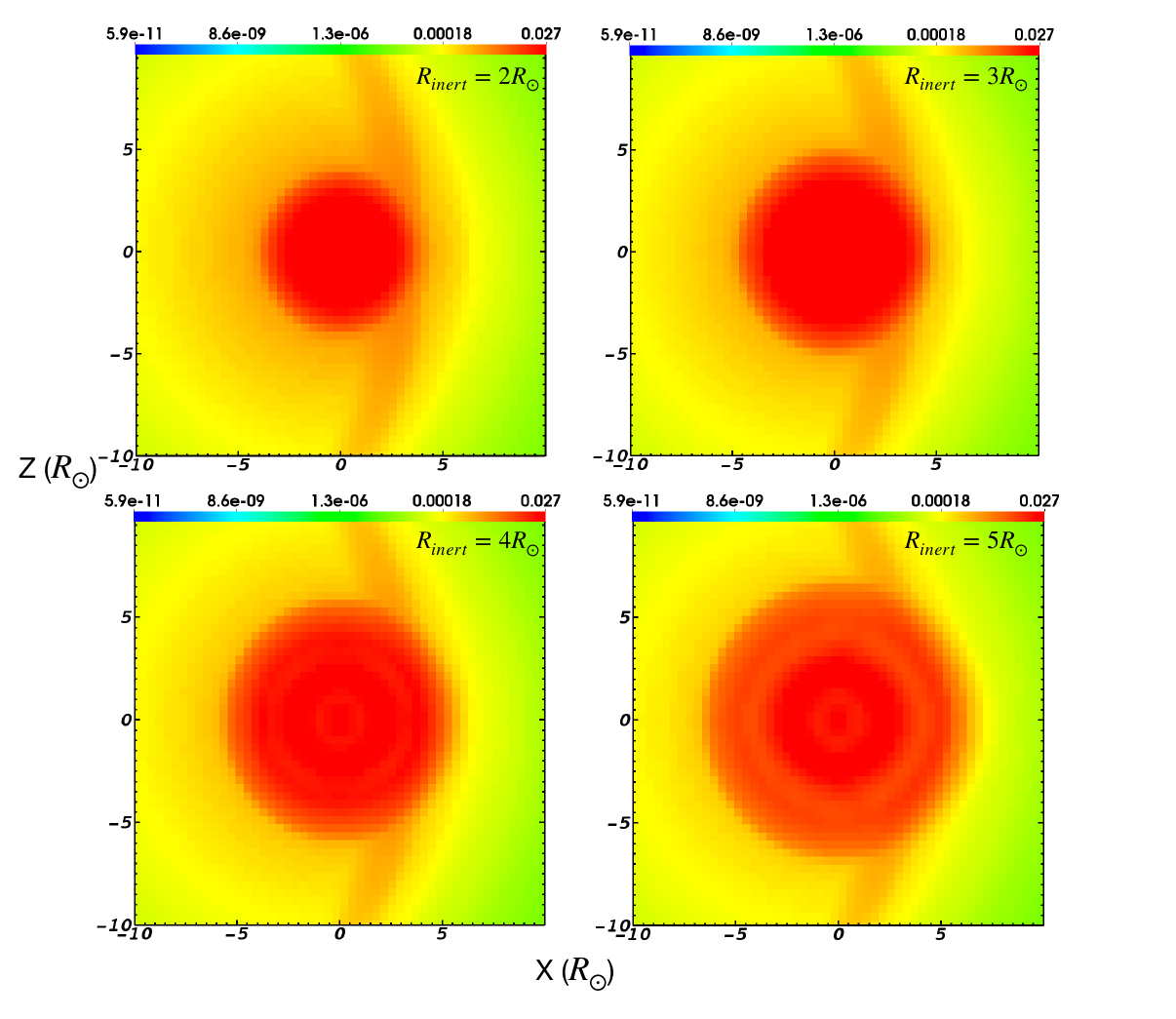} \\
\caption{Density maps in the inner $20 R_\odot$ x $20R_\odot$ at $t=1 \h$ for the four simulations with different values of $R_{\rm inert}=2,3,4, 5 R_\odot$ (simulations S1, S6, S7, S8). 
The colors depict the density values in $\g \cm^{-3}$ according to the color bar on the top of each panel. 
The red arc marks the shock wave that propagates from right to left. The central regions have densities above the maximum value of these plots of $0.027 \g \cm^{-3}$ and therefore they are coloured red. 
}
	\label{fig:one_h_Rinert}
\end{figure*}

The small increase of the leftover bound envelope mass with decreasing radius of the inert core might result from a larger dissipation of the shock energy when we include in the simulations the higher density zones around the core. In Fig. \ref{fig:one_h_Rinert} we present density maps in the same plane as in Fig. \ref{fig:dens_cont} but in the inner $20 R_\odot \times 20 R_\odot$ at $t=1 \h$ when the shock wave crosses the inner region and for the four cases with $R_{\rm inert}=2,3,4, 5 R_\odot$ (simulations S1, S6, S7, S8). We see that the size of the inert core does not influence much the shocks outside the inert core. This explains the small differences between the simulations that differ only by $R_{\rm inert}$.
% FFFFFFFFFFFFFFFFFFFFFFFFFFFFFFFFFFFFFFF

%FFFFFFFFFFFFFFFFFFFFFFFFFFFFFFF
    
When the merger-explosion energy and site are $(E_{\rm exp},X_{\rm mer})=(3 \times 10^{50} \erg, 50 R_\odot)$, namely, a lower merger-explosion energy, the final envelope mass is larger (dashed-black lines), and the shock does not remove mass from $r \la 20 R_\odot$. 
For the closer but the same merger-explosion energy case S4 (HR), $(E_{\rm exp},X_{\rm mer})=(10^{51} \erg, 25 R_\odot)$, the final leftover envelope mass is lower as expected (solid-cyan lines). 
In simulation S5 (HR) for which  $(E_{\rm exp},X_{\rm mer})=(3 \times 10^{50} \erg, 25 R_\odot)$ the leftover envelope mass (dashed-cyan lines) is similar to the case of $(E_{\rm exp},X_{\rm mer})=(10^{51} \erg, 50 R_\odot)$ (solid-magenta lines). 

Over all, most, $\ga 90\%$, of the envelope gas inner to the location of the merger site remains bound. This bound mass amounts to $\simeq 3.5-3.9 M_\odot$ for the stellar model we use. This falls within the general range of post-merger bound mass that \cite{Soker2021NSNS} estimated $M_{\rm PM,e} \approx 0.5-5 M_\odot$ (see section \ref{sec:Evolution}).

%===================================
\section{Discussion and Summary} 
\label{sec:Summary}
% ===========================================================

We simulated the influence that the energy that a NS-NS merger releases inside a RSG envelope has on the envelope inner to its orbit. We assumed that the merger sets a spherical explosion inside the envelope that excite a strong shock in the envelope (Fig. \ref{fig:SchematicScenario}). We set such an explosion at off-center location from the center of an inflated RSG model (Figs. \ref{fig:DensityProfile} and \ref{fig:Dens_mass}). In the upper left panel of Figs. \ref{fig:dens_cont}-\ref{fig:temp} we present the initial maps of density, of velocity, and of temperature for the S2 (HR) simulation (table \ref{Table:cases}). We present the evolution of these quantities in the later eight panels of  Figs. \ref{fig:dens_cont}-\ref{fig:temp}. 

From Fig. \ref{fig:dens_cont} we learn that the shock substantially deform the inner zones of the RSG at few hours after the merger-explosion. The shock deposits energy to the RSG and removes some envelope mass. Later, at about $t \simeq 10 \h$, the envelope near the center has relaxed for a while, as the more or less spherical inner density contour indicates in the lower right panel of Fig. \ref{fig:dens_cont}. 

Our main finding is that for typical NS-NS merger explosion energies and for mergers not too close to the center of the RSG, most of the envelope mass inner to the merger site, which is the orbit of the merger product, stays bound (Figs. \ref{fig:mass2040A} and \ref{fig:mass2040B}). This implies that the interaction of the merger product, a massive NS or a BH, with the envelope can release more energy \citep{Soker2021NSNS}. The merger product accretes more mass and launches jets that further power the transient event. As well, the leftover envelope causes further inspiralling of the merger product, a process that releases more orbital energy. For the parameters we chose here we do not expect the merger product to spiral-in all the way to the core. More massive RSG stars might maintain a more massive envelope and force the merger product to spiral-in to the core \citep{Soker2021NSNS}. 
    
The large fraction of triple stellar systems among massive stars motivated the study of massive triple star CEE scenarios (section \ref{sec:intro}). 
One such scenario is of a tight binary of two NSs that enters the envelope of an RSG and merges inside the envelope, which was the scenario we simulated here. 
The launching of jets by the two NS before merger and the merger inside the envelope lead to a very energetic transient event with total energy of $\ga 10^{52} \erg$ and a months-years long bright light curve \citep{Soker2021NSNS}. Other effects include the shortening of the time to NS-NS merger compared with that of isolated kilonovae. 

If the orbital plane of the tight NS-NS binary system is inclined to the orbital plane of the RSG with the tight binary, the final NS/BH-NS/BH binary (if bound) might have spin-orbit misalignment \citep{Soker2021AM}.   

\cite{Soker2021NSNS} estimates the ratio of NS-NS merger in CEJSN event rate to CCSNe rate as $\la 10^{-6} - 2 \times 10^{-5}$. However, because NS-NS merger in CEJSNe are much more luminous transients than typical CCSNe are their detection faction is higher. Our study shows that when the merger takes place in the envelope but not too close to the core the leftover envelope is large enough as not to terminate the powering of the CEJSN off.  Namely, the merger product can continue to accrete mass and launch jets, and even spirals-in into the core of the RSG star. This adds to the expectation that these events although very rare are also very bright. In the case of more massive RSGs that maintain a more massive leftover envelope after merger, the merger product might enter the core and launch jets that are a site of more r-process nucleosynthesis  \citep{GrichenerSoker2019a, Soker2021NSNS} in the early Universe. 

% ===================================================
\section*{Acknowledgments}
% ===================================================

We thank Aldana Grichener for supplying us the initial stellar model. 
We thank an anonymous referee for helpful suggestions and comments.
This research was supported by the Amnon Pazy Research Foundation.

%%%%%%%%%%%%%%%%%%%%%%%%%%%
\textbf{Data availability}
The data underlying this article will be shared on reasonable request to the corresponding author.  
%%%%%%%%%%%%%%%%%%%%%%%%%%%

% %%%%%%%%%%%%  References %%%%%%%%%%%%%%%%%%%%%

\label{lastpage}
\end{document}